\documentstyle[preprint,12pt,aps,psfig]{revtex}

\tighten
\begin{document}
\draft
\preprint{Imperial/TP/96-97/52}

\newcommand{\nc}{\newcommand}
\nc{\al}{\alpha}
\nc{\ga}{\gamma}
\nc{\de}{\delta}
\nc{\ep}{\epsilon}
\nc{\ze}{\zeta}
\nc{\et}{\eta}
\renewcommand{\th}{\theta}
\nc{\ka}{\kappa}
\nc{\la}{\lambda}
\nc{\rh}{\rho}
\nc{\si}{\sigma}
\nc{\ta}{\tau}
\nc{\up}{\upsilon}
\nc{\ph}{\phi}
\nc{\ch}{\chi}
\nc{\ps}{\psi}
\nc{\om}{\omega}
\nc{\Ga}{\Gamma}
\nc{\De}{\Delta}
\nc{\La}{\Lambda}
\nc{\Si}{\Sigma}
\nc{\Up}{\Upsilon}
\nc{\Ph}{\Phi}
\nc{\Ps}{\Psi}
\nc{\Om}{\Omega}
\nc{\ptl}{\partial}
\nc{\del}{\nabla}
\nc{\be}{\begin{eqnarray}}
\nc{\ee}{\end{eqnarray}}
\nc{\lambar}{\overline{\lambda}}

\title{Conformal Mapping, Power Corrections, and the QCD Bound State
       Spectrum}
\author{H. F. Jones$^a$\footnote{email: h.f.jones@ic.ac.uk}, 
        A. Ritz$^a$\footnote{email: a.ritz@ic.ac.uk}, and 
        I.L. Solovtsov$^{ab}$\footnote{email: solovtso@thsun1.jinr.dubna.su}\\ 
        $\;$\\}
\address{$^a$ Theoretical Physics Group, Blackett Laboratory, \\
         Imperial College, South. Kensington, London, SW7 2BZ, U.K.\\ $\;$\\}
\address{$^b$ Bogoliubov Laboratory of Theoretical Physics, 
         Joint Institute for Nuclear Research, \\
         Dubna, Moscow Region, 141980, Russia \\ $\;$}
\date{\today}

\maketitle

\begin{abstract}
We analyze the heavy quark bound state spectrum using an order-dependent
conformal mapping to re-sum the perturbative expansion for current correlators.
The procedure consists of two main steps. Firstly, the Borel plane structure
of the truncated perturbative expression is modified to 
ensure consistency with the operator product expansion. This is analogous
to a resummation of infrared renormalon chains. Secondly, this 
perturbative expansion is conformally mapped to a new series with
improved convergence properties. This approach may be shown to induce
power corrections consistent with existing condensates, and
the resulting expansion may be ordered in powers of an infrared-analytic
effective coupling.
The technique is then applied to
$c\overline{c}$ and $b\overline{b}$ sum rules without any explicit 
introduction of vacuum condensate parameters. 
Ground state masses for the vector, axial--vector and $A'$ channels
are well reproduced, while results for the
scalar--pseudoscalar mass splitting are less impressive. 
\end{abstract}


\section{Introduction}

The Wilson operator product expansion, within the framework of
QCD sum rules \cite{svz79a,svz79b,svz79c}
(see also \cite{rry85} for a review), constitutes the standard continuum
approach for the systematic treatment of nonperturbative corrections to 
perturbation theory in QCD, and study of the bound state resonances.
Taking asymptotic freedom and perturbative QCD as a starting point,
this formalism extends the validity of the operator 
product expansion (OPE) to the scale at which resonances appear in the spectrum
as a manifestation of confinement. This generalised operator product expansion 
for the correlators of various electromagnetic currents includes,
in addition to the perturbative contribution,
non-perturbative power corrections associated with vacuum condensates of 
the fields.

For a given current $j^{\Ga}$, we have
the correlator\footnote{We define $Q^2=-q^2$, with $q$ the 
current momentum.},
\be
 T_{\mu\nu\cdots}\Pi^{\Ga}(Q^2) & = & i\int d^4 x 
    e^{iqx} \left<0|T\{j^{\Ga}_{\mu\cdots}(x)j^{\Ga}_{\nu\cdots}(0)\}|0\right>,
\ee
where $T_{\mu\nu\cdots}$ is a tensor which depends on the Lorentz structure
of the particular current $j^{\Ga}$. The validity of the OPE
at the resonance scale requires the assumption that one may
separate the short distance (perturbative) and long distance (condensate)
contributions in a consistent manner. The expansion then has the form,
\be 
     i\int d^4 x e^{iqx} T(j^{\Ga}_{\mu\cdots}(x)j^{\Ga}_{\nu\cdots}(0)) 
      & = & C^{\Ga}_{pert}(Q^2){\rm {\bf 1}} + \sum_{n\geq 2} 
         \frac{C_W^{\Ga}(Q^2)}{Q^{2n}} {\cal O}_{2n},
\ee 
where $C^{\Ga}_{pert}(Q^2)$ is the perturbative contribution,
and ${\cal O}_{2n}$ are $2n$--dimensional gauge invariant operators
with $C_W^{\Ga}$ the corresponding perturbatively calculable 
Wilson coefficients. 

While the perturbative contribution is calculable in a consistent manner
in the short distance regime, provided one imposes an infrared (IR) cutoff,
the fact that the split is not necessarily unique or unambiguous is
signalled by the perturbative  
large order divergences associated with infrared 
renormalons \cite{thooft79,lautrup77,parisi78,parisi79,david84,mueller85}.
Indeed, it is well known that the Borel resummation ambiguity of the 
perturbative series for correlators associated with inclusive processes
has precisely the form, if not the magnitude \cite{dok96b}, 
of the power corrections determined by the non-perturbative condensate 
contribution to the OPE. The perturbative ambiguity may then be cancelled by 
a similar ambiguity in the condensates, rendering the combination well-defined.
Consequently, estimating the numerical value of condensates
requires detailed knowledge 
of the perturbative ambiguity itself, hence the recent interest in renormalon 
asymptotics (see e.g. \cite{zakharov92,beneke92,beneke93,grunberg94,beneke94,beneke95,ball95,bigi94,maxwell94,altarelli95,maxwell95,neubert95,neubert95c} and 
references therein).

Our aim in this paper is to consider the calculation of 
the mass of heavy quark bound states within the sum rules
formalism via an alternative approach which is to try to resum the asymptotic 
perturbative series into a convergent form. There are two clear 
problems that arise in this endeavour which need to be overcome: 
(1) The Borel non-summability
of the perturbative series, due to the presence of IR renormalons,
implies that such a resummation technique
must be powerful enough to induce power corrections to counteract
the fundamental ambiguity in the asymptotic perturbative series; 
(2) The Borel
resummation ambiguity, while having the correct functional form
to hint at the structure of the nonperturbative corrections required,
cannot itself provide the normalisation of these terms from within 
perturbation theory,
as the condensates have a non-perturbative origin associated with confinement
to which perturbation theory itself is ostensibly blind.

Nevertheless, with regard to point (1), order-dependent mapping techniques,
based on the resummation of a perturbative series in the coupling $g$,
$G(g)$, via a
conformal mapping 
\be
  g & \rightarrow &  \frac{a}{C(1-a)^{\al}}, \label{conf1}
\ee
to a new series in $a$, $G(g(a))$, \cite{seznec79,guida96} are powerful 
enough to resum Borel non-summable
series into convergent form. 
An example is the quantum mechanical double-well potential,
and we illustrate the structure of the induced corrections in a
$\ph^4_0$ model in Appendix A to indicate their connection
with field theoretic power corrections.  

The nonperturbative input stressed in point (2) can, however, only
arise in such an approach in a summation to all orders, and thus
cannot obviously be obtained with knowledge of only 
a finite truncation of the perturbative series, which
is the practical reality in field theoretic situations. Furthermore,
rigorous proofs of convergence require detailed knowledge of the
complex-analytic structure of $G(g)$, which is again generally lacking in 
$4D$ field theories. Thus, in this case one needs
extra information to try and ``optimise convergence'' of the conformally mapped
series. This is possible since in Eq.~(\ref{conf1}) there is a free
positive parameter $C$ ($\al$ is usually fixed via constraints on the mapping)
which may be fixed order by order to improve convergence. Generically,
a necessary condition is that $C$ must scale with the order of the expansion
in a specific manner. In QCD, with our knowledge limited to only the
lowest order terms, a natural way to achieve this is to use some 
infrared data related to confinement, as this
non-perturbative input allows us to estimate the appropriate scaling
of $C$ for the resummed series. Note, however, 
that our implicit parametrization
of the nonperturbative corrections which may be relevant is directly
related to the structure of the conformal mapping, and not just the
constant $C$ itself.

An order-dependent mapping approach of the kind discussed above
has been developed by Solovtsov \cite{solov94a,solov94b} 
(see also \cite{jones95a,solov95,jones95c,jones95b,jones96,jones97a}
for applications) which we briefly review in Section 2,
indicating how, at any finite order, the conformally mapped series
may be re-ordered into a form structurally equivalent to
perturbation theory, but with an infrared analytic effective coupling
which alters the IR behaviour of correlation functions.

For the application of this technique in situations where the numerical
value of condensate related power corrections is significant, and only
one- and two-loop perturbative coefficients are known, as in the sum-rules 
case, we also require a modification of the perturbative result to ensure the
correct all-order resummation ambiguity, i.e. the correct position
of the first IR renormalon pole. This then ensures consistency with the
OPE prior to resummation in the sense that the perturbative ambiguity 
may be consistently cancelled by ambiguities in the OPE condensates.
Or in the present context, that any induced power corrections compensating
the perturbative ambiguity can be associated with OPE condensates. 
An important point is that this modification is only necessary to
ensure the correct momentum dependence of the ambiguity, and thus
a full resummation is not required. The normalisation is provided
by the nonperturbative parameter $C$, and not the renormalon residue.

We illustrate this idea in the massless case in Section 2,
which was also discussed in \cite{jones97a}, 
while in Section 3 we apply the technique in the determination of
the masses of the $c\overline{c}$ and $b\overline{b}$ bound state families.
In this case we also require a resummation of threshold Coulomb singularities
and the particular resummation procedure adopted here is discussed in 
Appendix B. 
We then show that one may closely approximate the experimental estimates
in all channels other than the scalar and pseudoscalar. 
This suggests additional sources of power--behaved 
contributions in those channels to which this technique is insensitive. 
In Section 4, we discuss our results and consider the 
possible interpretations.

\section{An Order-Dependent Mapping in QCD}

In this section we briefly review how an order-dependent mapping
of Green functions may be obtained via direct manipulation of the
functional integral in QCD using the approach of Solovtsov
\cite{solov94a,solov94b} (see also \cite{jones95a,solov95,jones95c,jones95b,jones96,jones97a}), and then proceed to 
discuss the developments which facilitate later 
application to heavy quark sum rules.

At a formal level, the mapping may be implemented along the
ideas of Seznec and Zinn-Justin \cite{seznec79} via a direct
resummation of a perturbative series $G(....)=\sum_n c_n\la^n$ as an expansion
in a new parameter $a$, $G=G(\la(a))$, related to $\la=\al_s/(4\pi)$, by the
conformal mapping,
\be
  \lambda &\equiv& \frac{g^2}{{(4\pi)}^2} = \frac{1}{C}\,
         \frac{a^2}{{(1-a)}^3} ,  \label{lam}
\ee
where $C$ is a positive constant. Note that this is particular case
of the conformal mapping (\ref{conf1}) with $\al=3/2$, and one observes that
$0\leq a < 1$ for all values of the gauge coupling.

However, in the case of QCD, to ensure gauge invariance, and
to allow implementation of the renormalization group, it is convenient
to have an explicit realisation of this mapping. Such a realisation
has been developed by Solovtsov \cite{solov94a,solov94b},
and  becomes possible 
with the introduction of an auxiliary field $\chi^a_{\mu\nu}$ used to
split up the quartic gauge field interaction term, which we denote
$S_4(A)$\footnote{We note that the theory is still consistent off-shell. 
One may readily verify that if $\ch$ has a gauge transformation consistent
with its on-shell constraint, $\ch \sim gf AA$, then the full
effective action still satisfies the functional Slavnov-Taylor
identities.},
\be
 \exp(ig^2S_4(A)) & = & \int [d\ch] \exp\left[\frac{i}{2}\int dx \ch_{\mu\nu}^a
       \left(\ch^{\mu\nu a}+i\frac{g}{\sqrt{2}}f^{abc} 
         A^{\mu b}A^{\nu c}\right)\right]. \label{chi}
\ee
Use of this auxiliary field reduces all the interaction terms in the QCD action
to Yukawa form, and enables a new split between free and interaction
parts, $S_{\ch} = \tilde S_0 + \tilde S_I$, parametrised by 
the variables $\xi$ and $\ze$, where
\be
 \tilde S_0 & = & \ze^{-1}[S(A,\ch)+S_2(\ps)+S_2(c)]+\xi^{-1}S_2(\ch),
\ee
and
\be
 \tilde S_I & = & gS_3(A,\ps,c) - (\ze^{-1}-1)[S(A,\ch)+S_2(\ps)+S_2(c)]
   -(\xi^{-1}-1)S_2(\ch).
\ee 
We use a compact notation for the standard kinetic and Yukawa interaction
terms of the gauge ($A$), quark ($\ps$), and ghost ($c$) fields.
and there are also terms fixing the covariant $\al_G$ gauge.
$S_2(\ch)$ is a mass term for the $\ch$ field introduced in 
(\ref{chi}), while $S(A,\ch)$ is the gauge propagator in the
$\ch$ background (see \cite{solov94b}). A Green function for
an even number of fields then takes the form,
\be
 G(....) & = & \frac{1}{Z}\int [d\ch]
  [d(A,\ps,c)]\sum_{n=0}\frac{1}{n!}(....)(i\tilde S_I)^n 
       e^{i\tilde S_0},
\ee
where $[d(A,\ps,c)]$ denotes the conventional functional integral over
gauge, quark, and ghost fields, and $Z$ is the partition function.

As discussed in \cite{solov94a,solov94b}, further expansion 
of $\tilde S_I$, and a rescaling of the fields, allows $\ch$ to
integrated out, restoring the standard free action of QCD in the
exponential and the correct relationship between the 3-point and 4-point
gauge couplings, provided that $\xi=\ze^3$. However, on inspection of
the subsequent series, one observes that it may be ordered
in terms of a new parameter, $a\equiv 1-\ze$, {\it provided} one 
identifies $a$ with the gauge coupling precisely via the
conformal mapping (\ref{lam}). 

Expansion of $G$ as a power series in $a$, up to $O(a^5)$, then
results in the expression
\be
 G^{(5)}(..) & = & g_0(..) +\frac{a^2}{C}g_2(..) +3\frac{a^3}{C}g_2(..)
      +\frac{a^4}{C^2}[6Cg_2(..)+g_4(..)] \nonumber \\
      &  & \;\;\;\;\;\;\; +\frac{a^5}{C^2}[10Cg_2(..)+6g_4(..)] +O(a^6),
\ee
where $g_{2n}(..)$ are the corresponding terms in a standard
perturbative series with coefficient $\la^n$ \cite{solov94b}. 
Re-ordering this series in terms of the perturbative factors $g_{2n}(..)$
we find
\be
   G^{(5)}(..) & = & g_0(..) +\frac{a^2}{C}(1+3a+6a^2+10a^3+\cdots)g_2(..) 
        \nonumber \\
    & & \;\;\;\;\;\;\;+\frac{a^4}{C^2}(1+6a+\cdots)g_4(..)] +\cdots \\
   & = & g_0(..) + \la_{eff}|_5 g_2(..) +(\la_{eff})^2|_5 g_4(..) +\cdots,
\ee
where we have introduced the effective coupling, $\la_{eff}|_n$, which 
is the expansion of 
(\ref{lam}) to $O(a^n)$. Therefore, at a given finite order, the conformal
mapping may be realised by a direct replacement of the coupling $\la$
in the perturbative series with $\la_{eff}$ defined to the appropriate order
in $a$. We shall henceforth work at $O(a^3)$, and thus
\be
 \la_{eff} & = & \frac{a^2}{C} (1+3a). \label{lameff}
\ee

While the equation (\ref{lam}) implies a formally nonlinear relationship
between the power series in $\lambda$ and $a$, for the truncated series
truly non-perturbative input arises via the ability to fix the
variational parameter $C$ at each order. One thus obtains
a sequence of approximants for the quantity $G$ under study,
$\{G_1(O(a),C_1),G_2(O(a^2),C_2),\ldots,G_N(O(a^N),C_N)\}$, and the
choice of $C$ is dictated by the aim of optimising the
convergence of this sequence.

In the present case the normalisation of the result depends on
the value of condensates related to confinement effects which
are invisible at any finite order of perturbation theory.  
The structure of these corrections is implied by the 
large order ambiguity, but this does not assist in providing the
correct normalisation. Therefore we fix $C$ at each order 
by fitting a particular low-energy quantity, which ensures the correct
normalisation at that scale. The universality of this choice can
then by checked by applying the same method to different processes, and
the relative magnitude of successive terms in the series can still be 
monitored, even though a formal proof of convergence is 
lacking\footnote{Note that such series
have been proven to converge in simpler models 
\cite{buckley93,guida95,arvan95c}
for which the corresponding perturbative series is at best asymptotic,
and also as discussed in Appendix A in some situations where the
perturbative series is non-Borel-summable \cite{guida96}.}.

Given this philosophy, the approach which has been used (with previous 
success \cite{jones95a,solov95,jones95c,jones95b,jones96,jones97a})
is to ensure that the running coupling has a singular
infrared asymptotic behaviour consistent with the linear part of the static
quark potential, $V_{lin}(r) = \si r$. As discussed in \cite{solov94b}, 
this corresponds to the requirement that the $\beta$--function obey
$\beta(\la) \longrightarrow -\la$ for large $\la$.
Enforcement of this constraint requires us to consider the 
renormalisation group evolution of the parameters.
A convenient check on the gauge invariance of the expansion 
is provided by calculating
the charge renormalisation constant with an arbitrary covariant
gauge and ensuring that dependence on the gauge parameter drops
out at each order in $a$. This is indeed the case \cite{solov94b}, and to
$O(a^3)$ the $\beta$--function is given by \cite{solov94b}
\begin{eqnarray}
\label{beta}
 {\beta} (a) & = & -\frac{b_0}{C^2}\,
       \frac{a^4}{(2+a)\,(1-a)^2}\,(2+9\,a)\, ,
\end{eqnarray}
where $b_0$ is the first coefficient of the perturbative $\beta$--function.
The resulting RG equation has the form
\begin{eqnarray}
\label{rga}
 f(a) & = & f(a_0)+ \frac{2b_0}{C}\ln \frac{Q^2}{Q_0^2},
\end{eqnarray}
where at this order
\begin{eqnarray}
\label{f}
  f(a) & = & \frac{2}{a^2} -   \frac{6}{a} - 48\ln a-
    \frac{18}{11}\,\frac{1}{1-a} 
   +  \frac{624}{121}\,\ln{(1-a) } +
    \frac{5184}{121}\,\ln{( 1+\frac{9}{2}\,a )} \, .
\end{eqnarray}
The parameter $a_0$ and the momentum $Q_0$ in Eq.~(\ref{rga}) are 
defined at a
normalisation point for the effective coupling constant (\ref{lameff}).
Explicitly, using phenomenological data for the string tension $\si$, 
the above analysis of the $\beta$--function
has been shown to lead to $C=4.1$ \cite{solov94b} at this order. 
The running coupling is then given by ~(\ref{lameff}), 
with the running parameter $a(Q)$ determined implicitly by ~(\ref{rga}). 
As noted earlier, $a(Q^2)$ is finite at all scales and thus 
$\la_{eff}$ does not exhibit a Landau pole \cite{solov94b}.

Having fixed parameters in the infrared, 
the first consistency check is that we recover the perturbative 
expression for the running coupling at large energy scales. Indeed this is the
case. If, using (\ref{lameff}), (\ref{rga}), and (\ref{f}), we consider 
the regime $Q^2 \gg Q_0^2=\La_{QCD}^2$ where $\la \sim a^2/C \ll 1$, 
we recover the one--loop perturbative result,
$ \la(Q^2) \rightarrow 1/(b_0\ln(Q^2/\La_{QCD}^2))$.
The IR running of the coupling is, however, 
significantly modified \cite{solov94b}.
The running quark mass \cite{solov94a}, given by
\begin{eqnarray}
 m(Q^2) & = & m_0 \left(\frac{\lambda(Q^2)}{\lambda(Q_0^2)}
                   \right)^{\frac{\gamma_0}{b_0}}, \label{runm}
\end{eqnarray}
where $m_0=m(Q_0^2)$ and 
$\gamma_0=4$ is the first coefficient of the anomalous dimension, 
similarly coincides with perturbation theory at large scales
but has a modified IR behaviour.

We now consider the application of this approach to the current correlators
introduced in section~1. For illustration it is convenient to consider
first massless correlators \cite{jones97a}, and 
in particular the Adler $D$-function,
$D(Q^2)\equiv-Q^2$d$\Pi(Q^2)/$d$Q^2$.  
These arguments will then be extended to the
massive case in Section 3. 

We begin with the standard spectral representation,
\be
\label{spec}
 D(Q^2,\la) & = & Q^2 \int_0^{\infty} {\rm d}s \frac{1}{(s+Q^2)^2}R(s,\la),
\ee
where $R(s)\equiv Im\Pi(s+i\ep)/\pi$ is given in a convenient
normalisation at $O(\la)$ by
\be
 R(s) & = & 1+4 \la.
\ee
Conventional RG improvement of the $D$-function would break its analytic 
properties by introducing the perturbative Landau pole at $Q^2=\La_{QCD}^2$.
However, this may be circumvented within the spectral representation
using RG improvement of the integrand \cite{ginzburg66}, with the
knowledge that $R(s)$ obeys the same homogeneous RG equation as $D$. This
corresponds to choosing a more general solution of the RG equation
for $D$ with the same UV asymptotics.
However, the integral in (\ref{spec}) is then undefined, 
as it now runs over the
Landau pole. The structure of the virtuality distribution in 
(\ref{spec}) implies that there is an infrared renormalon ambiguity
of $O(\La^2_{QCD}/Q^2)$. However, there is no corresponding
condensate operator in the OPE, and thus this first IR pole is expected
to be absent, at least within the current perturbative analysis. 
An all-order resummation in the large $b_0$ limit
would indeed recover the expected branch structure, removing the
first IR renormalon pole. 

We shall make the assumption that the first IR pole should be absent
from the point of view of perturbative asymptotics, to ensure
consistency with the OPE\footnote{Note that
this excludes the possibility that such a correction could arise as an
exponentially suppressed correction to the coefficient of the identity
operator in the OPE.}. However, as discussed in
Section 1, a full resummation is unnecessary in the present context, as the 
normalisation is fixed elsewhere. Thus we can use a simple trick
to remove the first IR pole and ensure that the large order
asymptotics are consistent with the OPE. A convenient way to achieve this
is to perform an integration by parts in (\ref{spec}) \cite{jones97a},
obtaining
\be
 D(Q^2) & = & 2Q^2 \int \frac{s{\rm d}s}{(s+Q^2)^3} R(s) -Q^2 \frac{{\rm d}}
 {{\rm d}Q^2}D(Q^2). 
\ee

\begin{figure}
 \centerline{%
   \psfig{file=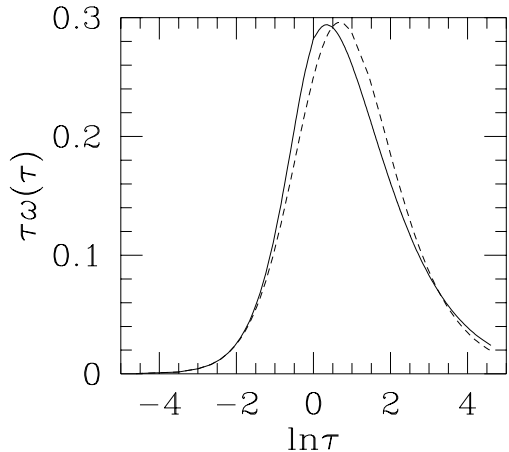,width=7cm,angle=0}%
   }
 \vspace*{0.2in}
 \caption{The virtuality distribution functions
  $\ta\,\om (\ta)$ taken from Ref.~\protect\cite{neubert95} 
  (solid line) and the
  function (\protect\ref{virtdis}) multiplied by a factor of 
  $\ta$ (dashed line) versus
  $\ln \tau$.}
        \label{virt}
\end{figure}

\noindent At the order 
to which we are working the second term vanishes, and we find
\be
 D(t,\la) & = & 1+4\int_0^{\infty}{\rm d}\ta \om(\ta) \la (\mu^2),
    \label{spec2}
\ee
where $t=Q^2/\mu^2$. The weight function, given by
\be
 \om(\ta) & = & \frac{2\ta}{(1+\ta)^3}, \label{virtdis}
\ee
describes the distribution of virtuality which we shall comment on shortly.
Performing RG improvement on the integrand, using for consistency
the one--loop $\beta$--function $\beta(\la)=-b_0\la^2$, we obtain
\be
 D(t,\la) & = & 1+4\int_0^{\infty}{\rm d}\ta \om(\ta)
                   \frac{\overline{\la}}{1+\overline{\la}b_0\ln\ta},
  \label{rgspec}
\ee
where $\overline{\la}=\overline{\la}(t)$.\footnote{We denote perturbative
running parameters in the $\overline{MS}$ scheme with a bar, as opposed to
the unbarred running parameters in the variational series.}
The modified structure of the virtuality distribution, which at low
scales now has a linear dependence on $\ta$, $\om(\ta)\sim 2\ta +O(\ta^2)$, 
is now consistent
with an ambiguity associated with the first IR renormalon pole
of $O(\La^4_{QCD}/Q^4)$ which can be consistently
cancelled by an associated ambiguity
in the gluon condensate in the OPE.

We observe that the 
virtuality distribution function (\ref{virtdis}) coincides
with the function used in \cite{bigi94} and remarkably 
is numerically very close to that
obtained in \cite{neubert95} (see Figure 1) which, in contrast
to the present naive modification, corresponds to an all orders 
resummation of renormalon contributions in the large $b_0$ limit.

This connection with renormalons is illustrated more clearly by 
performing a formal Borel transformation on (\ref{rgspec}), whereby
we obtain
\be
  D(t,\la) & = & 1+4\int_0^{\infty} {\rm d}b 
      \exp\left(-\frac{b}{\lambar(t,\la)}\right)B(b),
\ee
with
\be
 B(b) & = & \Ga(1+bb_0)\Ga(2-bb_0).
\ee
This Borel function exhibits the correct infrared and 
ultraviolet renormalon poles for the $D$--function, although not the full
branch structure \cite{mueller85}. Nonetheless, as mentioned above,
the crucial point here in using the partial integration in (\ref{virtdis})
is to obtain the correct positioning of the poles
and thus an ambiguity consistent with the OPE.

Having massaged the perturbative series into a form whose
large order divergence may be consistently cancelled by
ambiguities in the OPE condensates we may now perform the
conformal mapping, as discussed earlier, by replacing the 
perturbative running coupling with $\la_{eff}$ (\ref{lameff}).
In the modified spectral representation, this leads to the replacement of 
(\ref{rgspec}) by
\be
 \label{npspec}
 D(t,\la) & = & 1+4\int_0^{\infty}{\rm d}\ta \om(\ta) \la_{eff}(t\ta),
\ee
where, to $O(a^3)$, $\la_{eff}$ is given by (\ref{lameff}), with the
running expansion parameter $a=a(Q^2)$ determined 
via (\ref{rga}).

Within a field theory such as QCD it is 
not possible to directly assess the convergence
of this series, and therefore to determine whether indeed  OPE type
power corrections of the appropriate magnitude are induced when compared
to the initial perturbative series\footnote{Note that a $1/Q^2$ 
correction may also be induced in the
running coupling by removal of the Landau pole \cite{ball95}. However,
this has a purely perturbative short distance origin. For recent work
see \cite{grunberg97,shirkov97}.}. Although analysis of a $\ph^4_0$ model
\cite{guida96} (see Appendix A) indicates that the mapping is able to 
resum the appropriate corrections, within QCD it is not
always possible to assess even the magnitude of higher order terms.
Therefore we propose instead to test the technique by analysing the massive
case, and comparing the results directly with experiment.

\section{Moments for Heavy $q\overline{q}$ Bound States}

We shall make use of the standard sum rules approach for the 
determination of the lowest mass 
resonances\cite{svz79a,svz79b,svz79c} (see also 
\cite{rry85} for a review), wherein one assumes the validity of the
narrow resonance approximation for the lowest mass contribution to
the imaginary part of the two--point current correlator. The other side of the
sum rule is conventionally
determined directly from QCD up to parameters associated
with vacuum condensate operators. However, in the present 
context, as elaborated earlier, there is no distinct split between
perturbative and nonperturbative contributions. The non-perturbative
parameter $C$ does not have any direct connection with vacuum
condensates.

We now repeat the analysis of the preceding section in the case of
massive correlators.
To $O(\la)$ in perturbation theory we have
\begin{eqnarray}
 Im \Pi^{\Gamma}(s) & = &  \frac{1}{4\pi} \left[\Pi_{\Gamma}^{(0)}(s)
         +4\lambda  \Pi_{\Gamma}^{(1)}(s)\right],
\end{eqnarray}
where $\Gamma$ denotes the current in question, 
and $s=q^2$. The one- ($\Pi^{(0)}$) and
two--loop ($\Pi^{(1)}$) components have been obtained previously and 
are enumerated in Appendix B to fix our notation. The first convergent 
moment is defined by \cite{rry85}
\begin{eqnarray}
 M_{1+N_{\Gamma}}^{(\Gamma)}(Q^2) & \equiv & -\frac{d\Pi^{\Gamma}(Q^2)}{dQ^2}
                 \nonumber\\
        & = & \frac{1}{\pi} \int_{4m^2}^{\infty} \frac{ds}{(s+Q^2)^2} Im 
                   \Pi^{\Gamma}(s),
\end{eqnarray}
which is also the Adler $D$-function up to a factor of $Q^2$. Introducing
$\sigma=s-4m^2$, which we associate with the virtuality, and the quantity
$u^2=\sigma/(\sigma+4m^2)$, we have
\begin{eqnarray}
 &&M_{1+N_{\Gamma}}^{(\Gamma)}(Q^2) =  \frac{1}{4\pi^2} \int_0^{\infty}
    d\sigma
     \frac{(\sigma+4m^2)^{N_{\Gamma}}\left(\Pi_{\Gamma}^{(0)}(u)
      +4\lambda \Pi_{\Gamma}^{(1)}(u)\right)}
      {(Q^2+\sigma+4m^2)^{2+N_{\Gamma}}}. \label{mom1}
\end{eqnarray}

An additional problem encountered in studying moments for
heavy $q\overline{q}$ systems is the dominant Coulomb interaction, which
leads to the perturbative series having an effective expansion
parameter given by $\la/u$. In the conventional approach, higher order
moments are dominated by the nonrelativistic region of small $u$
and thus the perturbative expansion breaks down due to the size
of the expansion parameter \cite{coulomb}. 
Since our approach still makes use of these
perturbative coefficients a resummation is required in order to
consider the moments for large $n$.

When, in analogy with the massless case, we perform an integration by parts
to remove the first IR renormalon pole, a particular resummation
which places higher order corrections inside the virtuality 
distribution becomes very natural. This resummation may be performed
exactly, and in Appendix C we describe its implementation in a simplified
model illustrating its effectiveness in resumming Coulomb singularities,
and its connection with the familiar Sommerfeld-Sakharov 
factor \cite{SSfactor}.

Implementation of the resummation in the present case
results in the following expression:
\begin{eqnarray}
 M_{1+N_{\Gamma}}^{(\Gamma)}(Q^2) & = & \frac{1}{4\pi^2} \int_0^{\infty}
    d\sigma \frac{(\sigma+4m^2)^{N_{\Gamma}}\Pi_{\Gamma}^{(0)}(u)}
       {(Q^2+W(\sigma))^{2+N_{\Gamma}}}, \label{m1}
\end{eqnarray}
where to $O(\la)$
\begin{eqnarray}
 W(\sigma) & = & (\sigma+4m^2)\left(1-4\lambda\frac{\psi_{\Gamma}(u)}
                      {\Pi_{\Gamma}^{(0)}(u)}\right),\label{ws}
\end{eqnarray}
and
\begin{eqnarray}
 \psi_{\Gamma}(u) & = & (1-u^2)^{1+N_{\Gamma}} \int_0^u du \frac{2u}
           {(1-u^2)^{2+N_{\Gamma}}} \Pi_{\Gamma}^{(1)}(u).
\end{eqnarray}

Carrying out renormalisation group improvement of the 
integrand\footnote{Note that for currents with a non-zero 
anomalous dimension the modification takes the form of 
an overall factor in the integral which will not contribute
to the moment ratios for large $n$, and thus not to the 
asymptotic estimates for the bound state masses. For this reason 
we may ignore any anomalous dimensions.},
the solution of the RG equation corresponds to 
(\ref{m1}) with $\lambda$ replaced by 
$\overline\lambda(\sigma)$ and $m$ replaced
by $\overline{m}(\sigma)$, where in the $\overline{MS}$ scheme the 
arguments of these running parameters are scaled by $k_{\lambda}=\exp(-5/3)$
\cite{neubert95b}. 

The conformal mapping is then realised via the replacement
of perturbative running parameters with the
non-perturbative running parameters in (\ref{lameff}) and (\ref{runm}),
and in place of (\ref{ws}) we have
\begin{eqnarray}
 W(\sigma) & = & (\sigma+4m^2(k\sigma))\left(1
      -4\lambda_{eff}(k\sigma)\frac{\psi_{\Gamma}(u)}
                      {\Pi_{\Gamma}^{(0)}(u)}\right),\label{wsrg}
\end{eqnarray}
where now $u^2=\sigma/(\sigma+4m^2(k\sigma))$.

Finally, we may obtain the power moments for arbitrary $n$, 
which are given by
\begin{eqnarray}
 M_n^{(\Gamma)}(Q^2) & = & \frac{1}{4\pi^2} \int_0^{\infty}
    d\sigma \frac{(\sigma+4m^2(k\si))^{N_{\Gamma}}\Pi_{\Gamma}^{(0)}(u)}
       {(Q^2+W(\sigma))^{1+n}}.
\end{eqnarray}
As in the standard sum rules approach, the mass of the first resonance
is then obtained by considering ratios of the moments,
$R_n^{\Gamma}=M_{n-1}^{\Gamma}/M_n^{\Gamma}$, for large $n$.

Using the perturbative formulae for the imaginary parts of the two-point
correlators listed in Appendix B the integral for the moments can 
be evaluated numerically for each current.
There are only three parameters in this calculation:
a reference value for the $QCD$ coupling $\alpha_0$, taken at the $\tau$
mass scale; the quark mass $m_c$ (or $m_b$); and the variational
parameter $C$, constrained to be near 4.1 via data from the quark-antiquark
potential. We have previously presented
an analysis of $\tau$--decay using this technique
\cite{jones97a} which resulted in a particular normalisation
of the coupling at the $\tau$ scale, $\alpha_0(M_{\tau})$. 
For consistency we use the
value extracted at $O(a^3)$, $\alpha_0=0.379$, 
which corresponds to the order of the
perturbative expressions with which we are working. 

An important aspect of the modification of the virtuality distribution
(Fig.~1) invoked to ensure consistency with the OPE, is that the
peak is shifted to higher scales. Consequently, implementation
of the Coulomb resummation produces a saddle point for the moments
which dominates the expression for large $n$ (see Appendix C). 
The existence of the  saddle point is a result of the
function $W(\sigma)$
attaining a minimum, at say $\sigma=\tilde\sigma$. This scale
$\tilde\sigma$ then dominates the expression for the moment as $n$
becomes large. Analysis of the various channels indicates
that this minimum occurs in the range $\tilde\sigma\sim 4-6$ GeV$^2$ 
for $c\overline{c}$, and $\tilde\sigma\sim 20-30$ GeV$^2$
for $b\overline{b}$. The fact that the saddle point occurs well
above $\Lambda_{QCD}$ provides additional justification for the
validity of the perturbative expressions with which we are working.
It is then clear that for large $n$, the moment ratio tends asymptotically 
to its saddle point approximation, i.e.
\begin{eqnarray}
 R_n(Q^2) & \stackrel{n\rightarrow\infty}{\longrightarrow} & 
        Q^2 + W(\tilde\sigma).
\end{eqnarray} 
The corresponding 
ratio, $R_n^{{\rm had}}(Q^2)$, which arises from assuming the 
validity of the narrow resonance hypothesis, has the form $Q^2+M_R^2$ for
large $n$, where $M_R$ is the mass of the first resonance in the 
relevant channel. Thus we obtain the prediction
\begin{eqnarray}
 M_R & = & \sqrt{W(\tilde\sigma)},
\end{eqnarray}
and as a consequence there is no requirement
to fix $Q^2$ explicitly, for example at $Q^2=0$,
in order to determine the mass of the 
resonance.

\begin{figure}
 \centerline{%
   \psfig{file=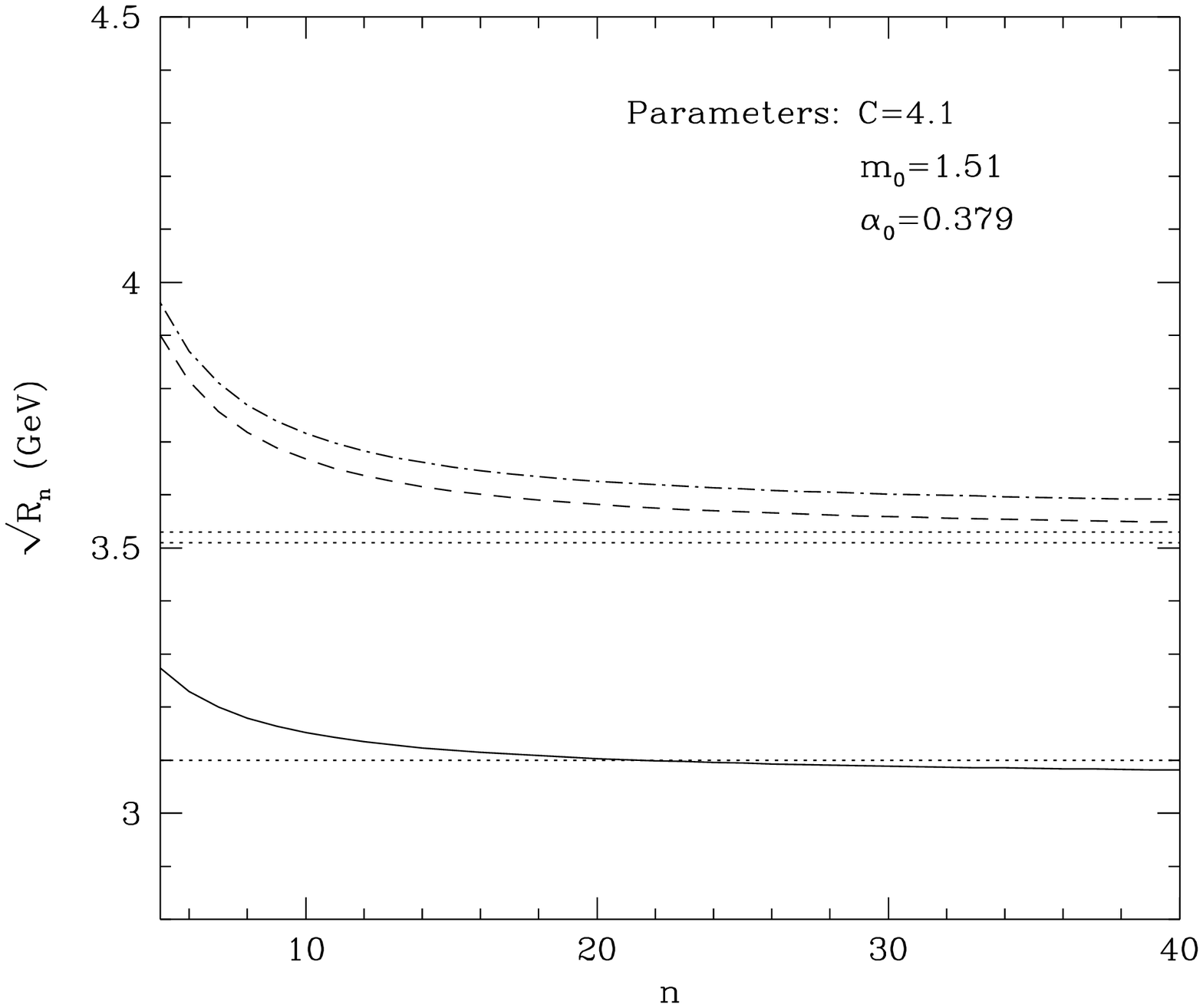,width=8.0cm,angle=0}
   \psfig{file=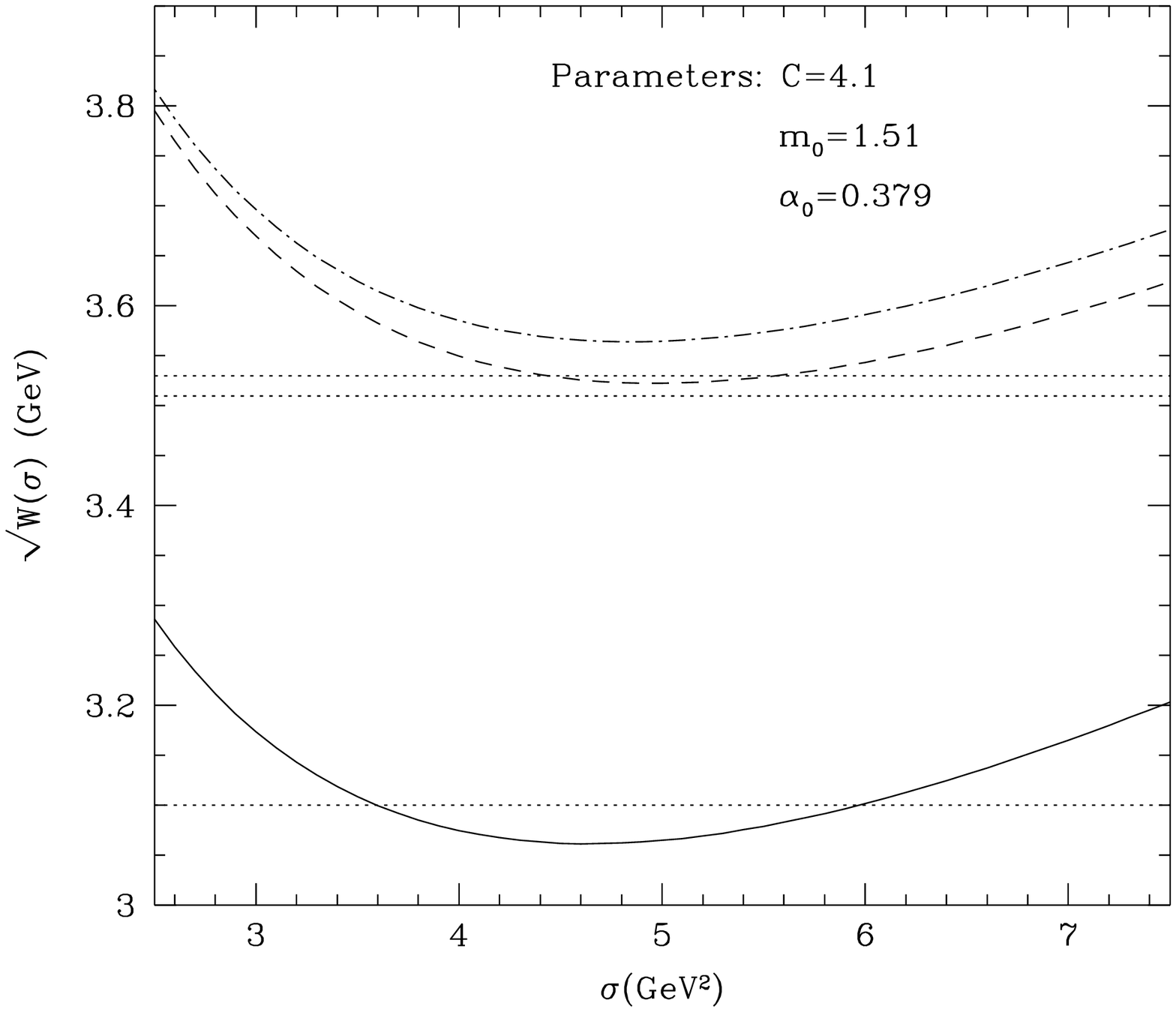,width=8.0cm,angle=0}%
   }
 \caption{On the left we plot ratios of the vector (solid curve),
  axial vector (dashed curve) and $A'$ (dot-dashed curve) moments for 
  $c\overline{c}$ bound states, and on the right we plot 
  $\protect\sqrt{W(\sigma)}$ versus $\sigma$ for the same currents,
  the minimum being the
  asymptotic limit of the moment ratios for large $n$. For comparison,
  the straight lines are, in order of decreasing mass, the corresponding
  experimental $c\overline{c}$ $A'$, axial-vector, and vector 
  bound state masses \protect\cite{pdg96}.} 
     \label{ccvaa}
\end{figure}

\begin{figure}
 \centerline{%
   \psfig{file=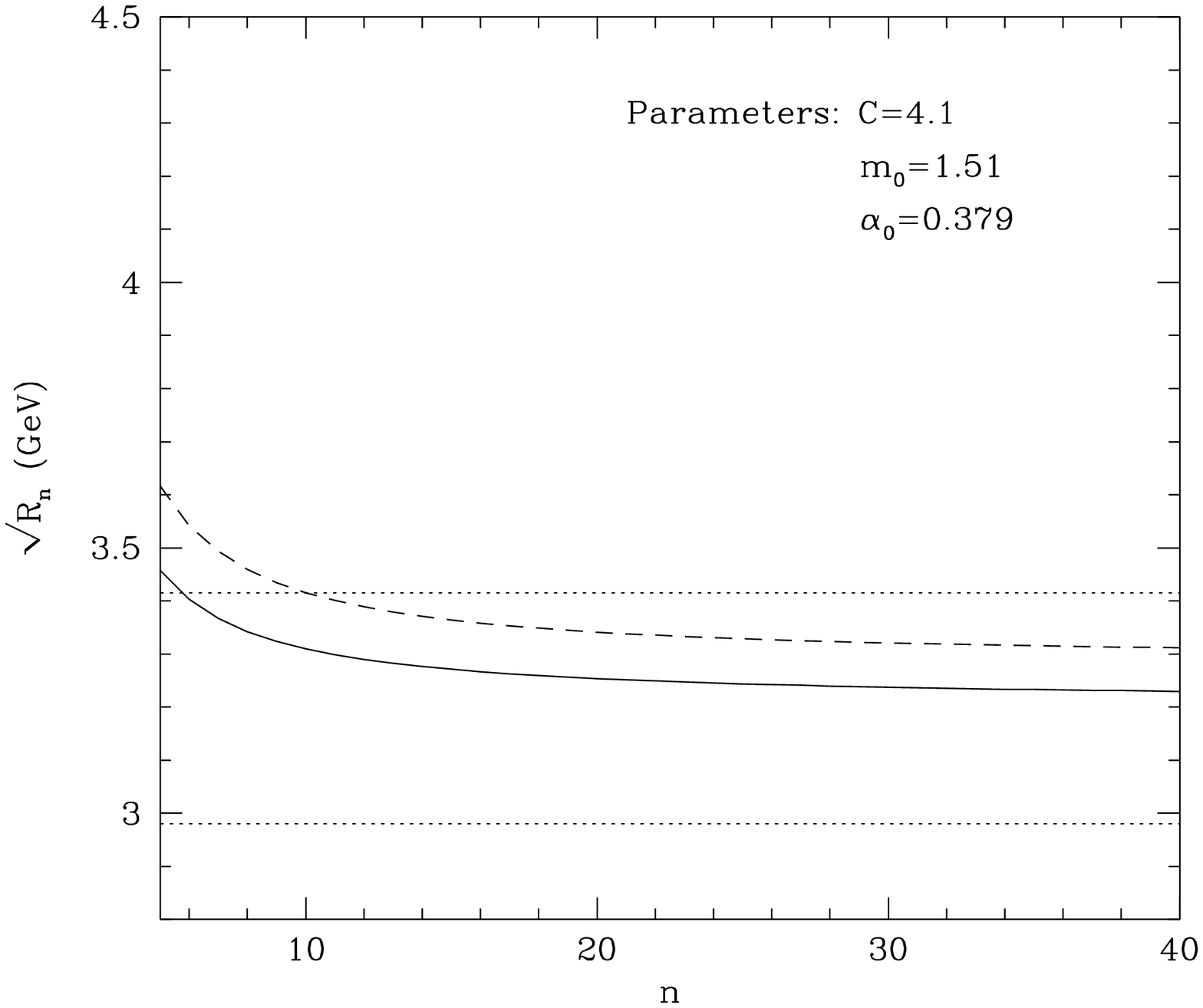,width=8.0cm,angle=0}
   \psfig{file=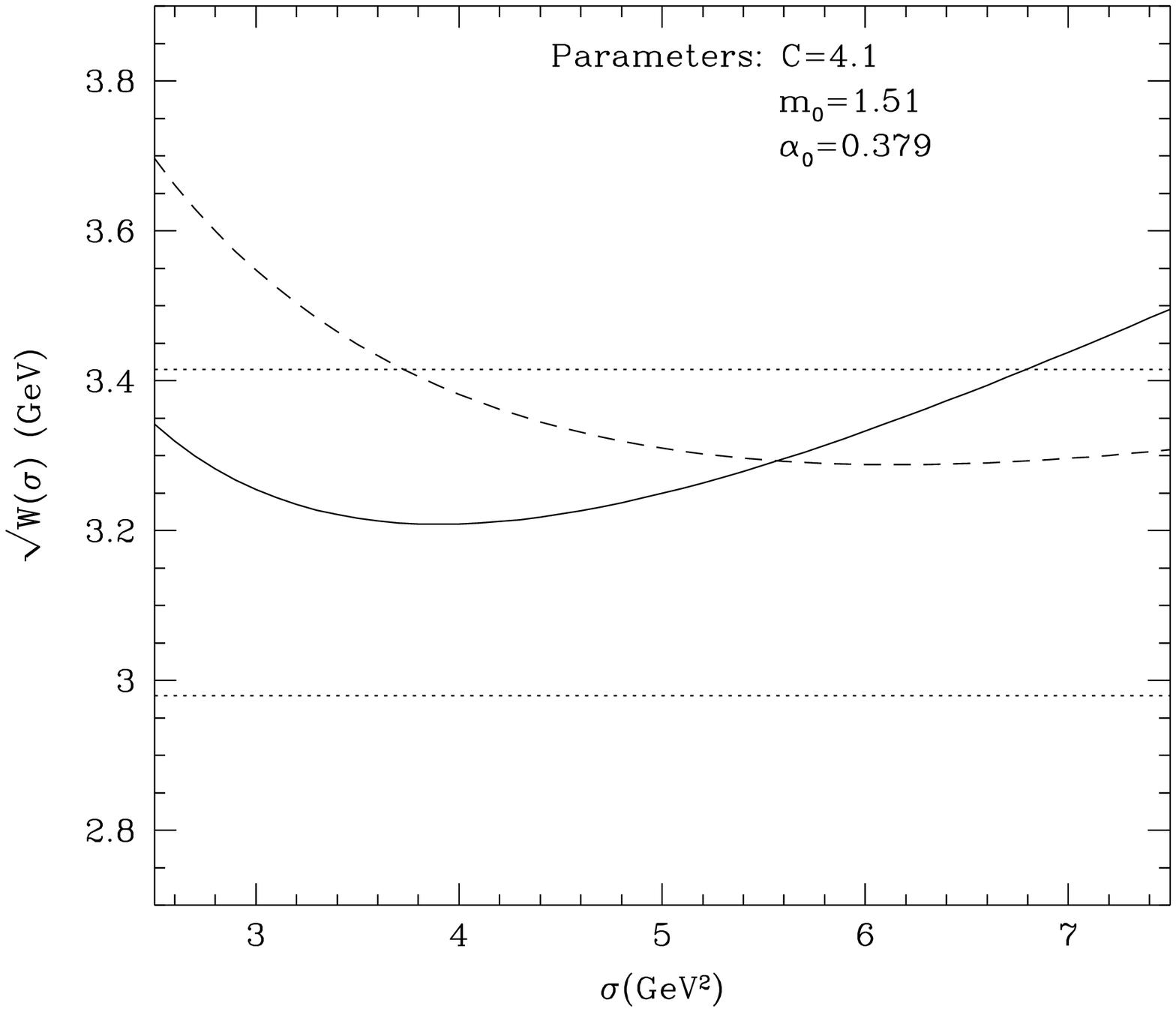,width=8.0cm,angle=0}%
   }
 \caption{On the left we plot ratios of the pseudo-scalar (solid curve), and
  scalar (dashed curve) moments for 
  $c\overline{c}$ bound states, and on the right we plot 
  $\protect\sqrt{W(\sigma)}$ versus $\sigma$ for the same currents,
  the minimum being the
  asymptotic limit of the moment ratios for large $n$. For comparison,
  the straight lines are, in order of decreasing mass, the corresponding
  experimental $c\overline{c}$ scalar and pseudo-scalar
  bound state masses \protect\cite{pdg96}.} 
     \label{ccsp}
\end{figure}

\begin{figure}
 \centerline{%
   \psfig{file=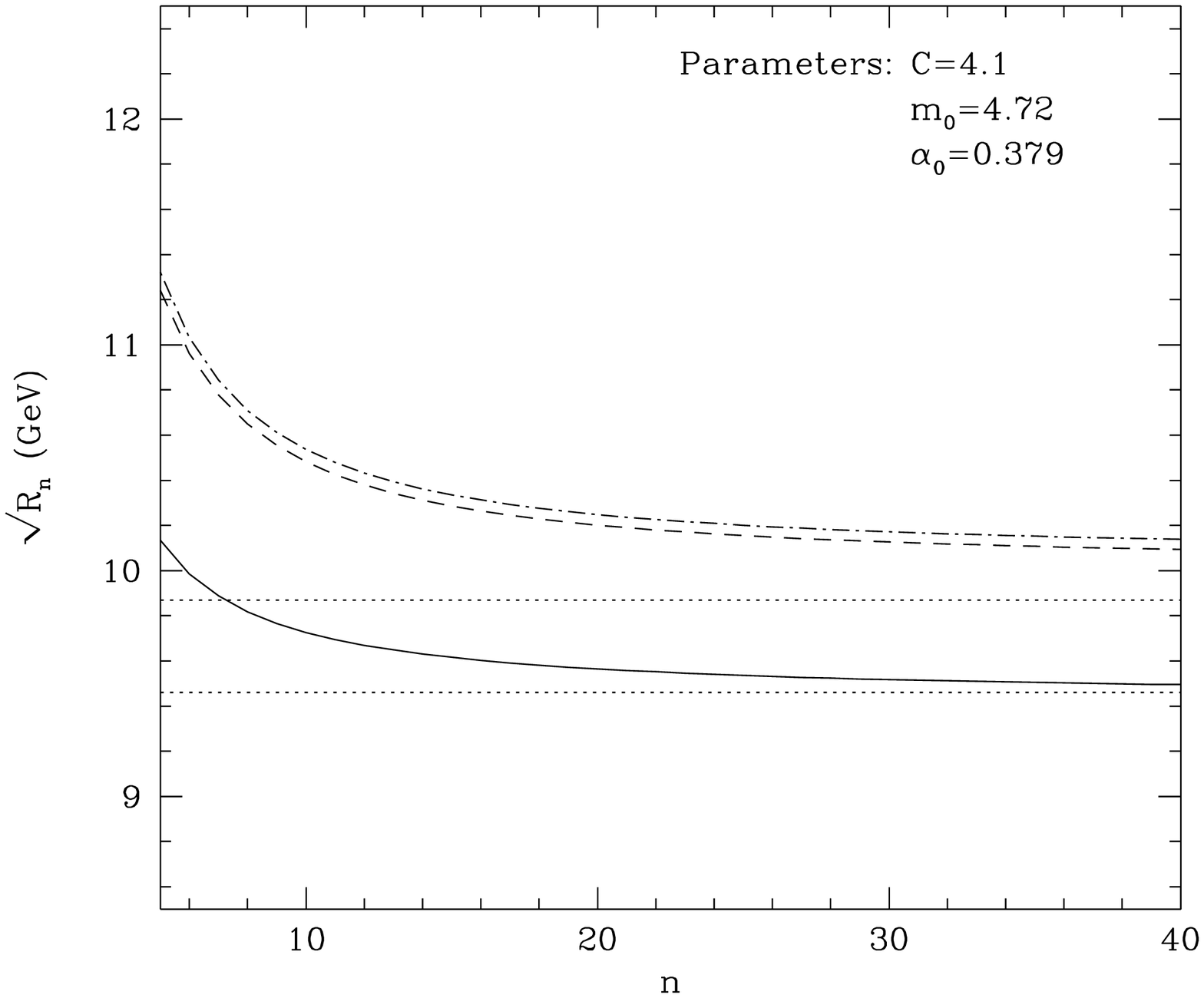,width=8.0cm,angle=0}
   \psfig{file=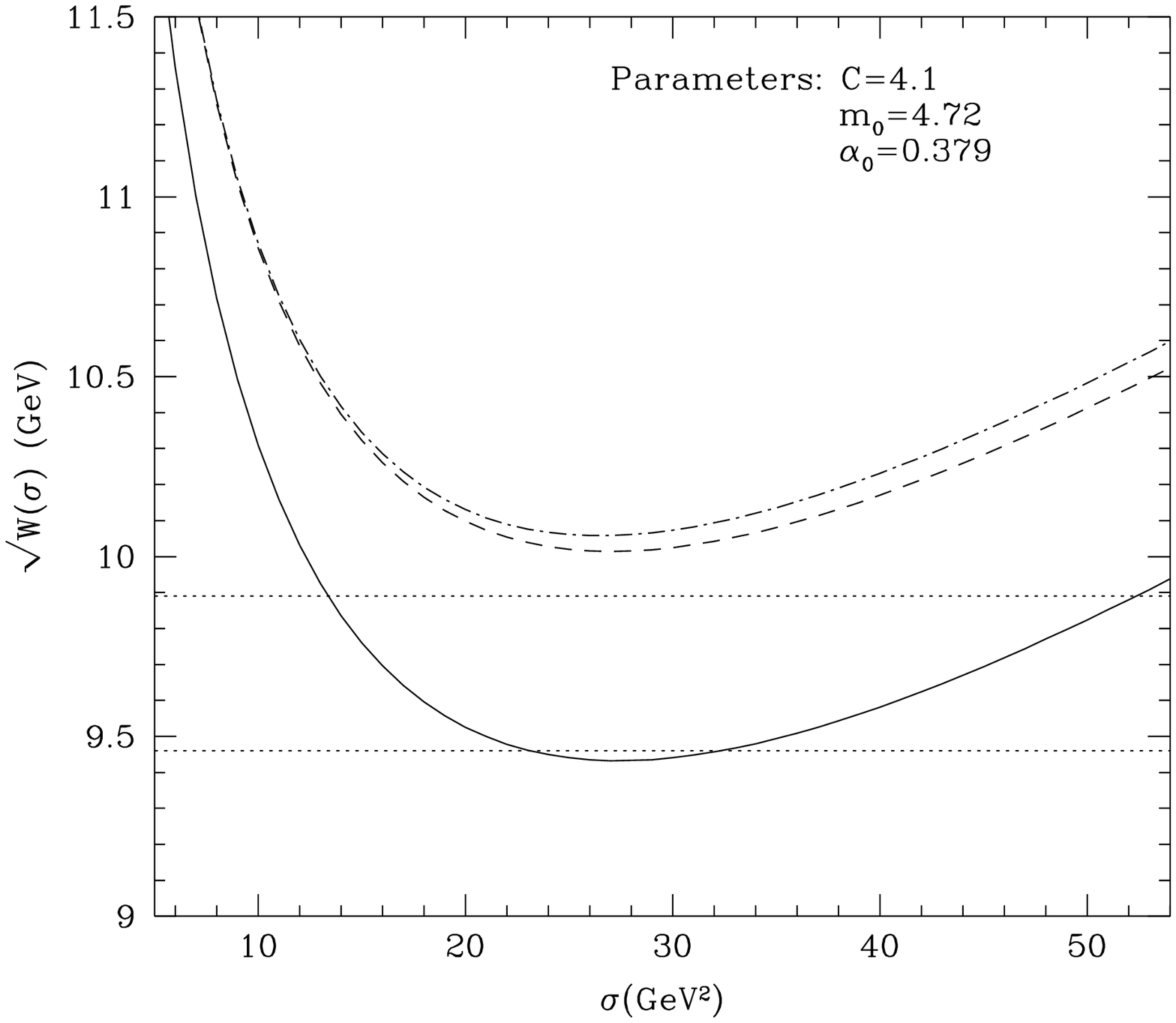,width=8.0cm,angle=0}%
   }
 \caption{On the left we plot ratios of the vector (solid curve),
  axial vector (dashed curve) and $A'$ (dot-dashed curve) moments for 
  $b\overline{b}$ bound states, and on the right we plot 
  $\protect\sqrt{W(\sigma)}$ versus $\sigma$ for the same currents,
  the minimum being the
  asymptotic limit of the moment ratios for large $n$. For comparison,
  the straight lines are, in order of decreasing mass, the corresponding
  experimental $b\overline{b}$ bound state masses 
  for the axial-vector and vector channels \protect\cite{pdg96}.} 
     \label{bbvaa}
\end{figure}

\begin{figure}
 \centerline{%
   \psfig{file=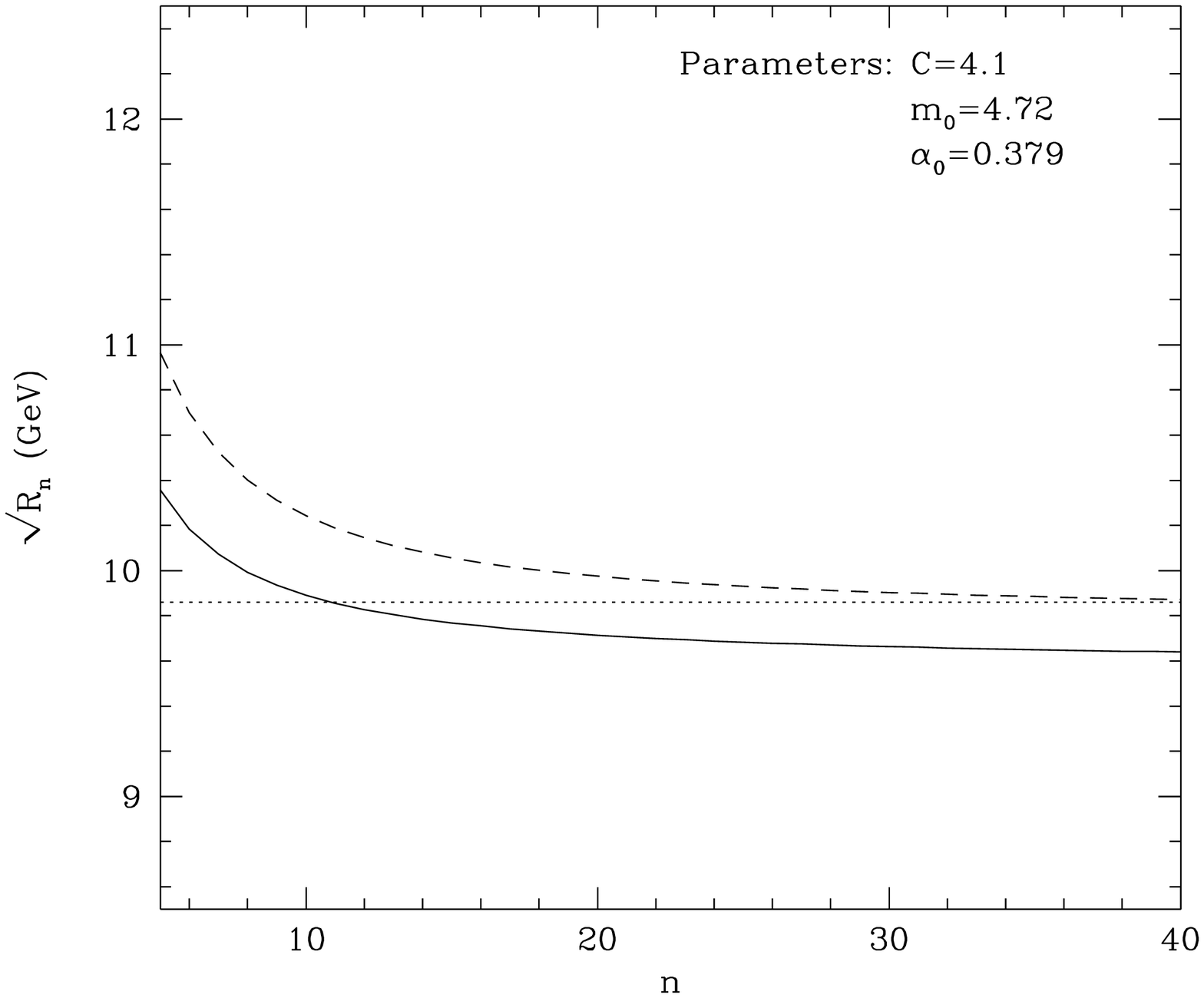,width=8.0cm,angle=0}
   \psfig{file=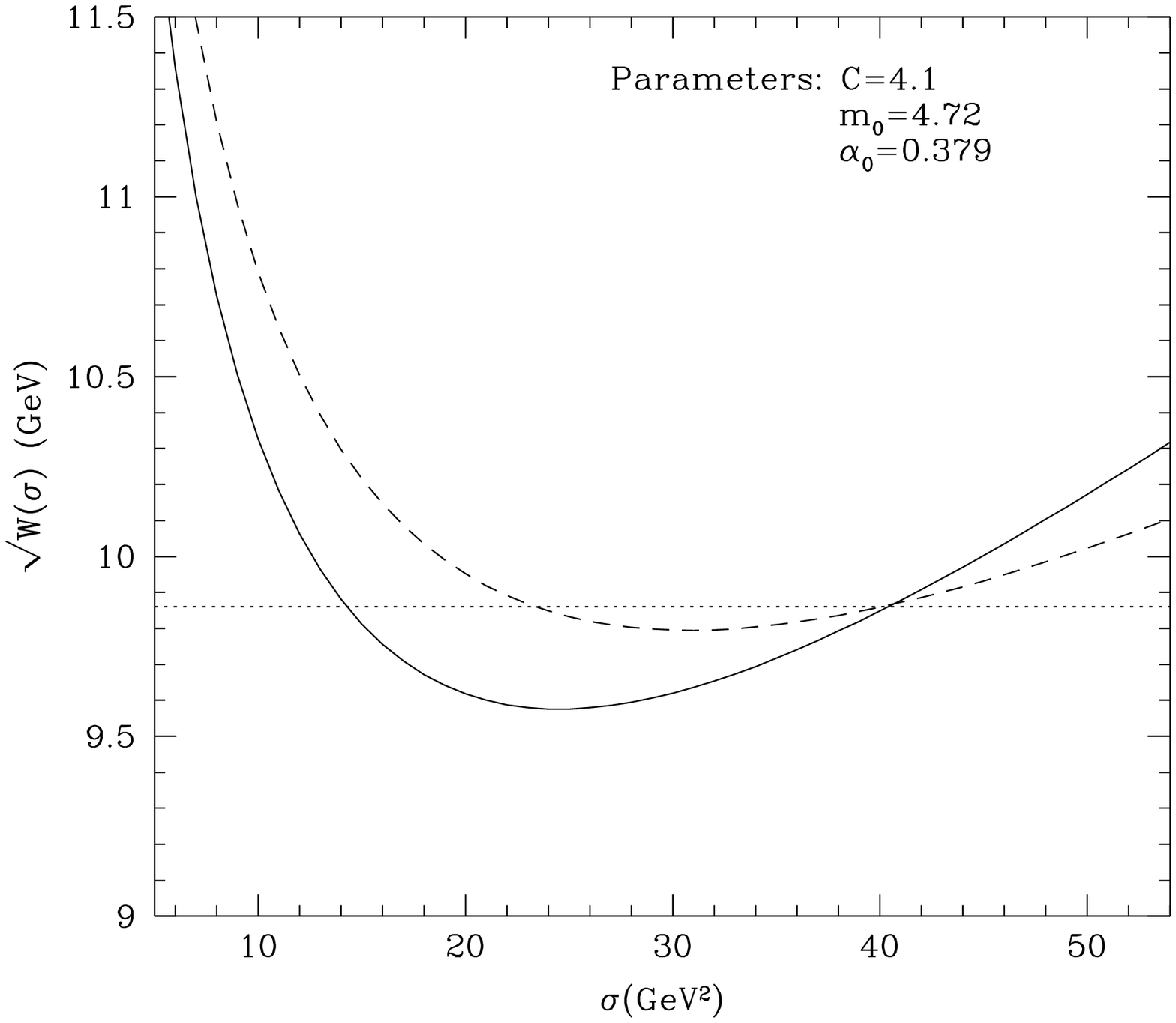,width=8.0cm,angle=0}%
   }
 \caption{On the left we plot ratios of the pseudo-scalar (solid curve), and
  scalar (dashed curve) moments for 
  $b\overline{b}$ bound states, and on the right we plot 
  $\protect\sqrt{W(\sigma)}$ versus $\sigma$ for the same currents,
  the minimum being the
  asymptotic limit of the moment ratios for large $n$. For comparison,
  the straight line is the corresponding
  experimental scalar $b\overline{b}$ bound state mass \protect\cite{pdg96}.} 
     \label{bbsp}
\end{figure}

Results of the numerical calculations for the full moment ratios, and the
asymptotic estimates $\sqrt{W(\tilde\sigma)}$, 
for all currents under consideration and for both
$c\overline{c}$ and $b\overline{b}$ bound states are shown in Figures
\ref{ccvaa}--\ref{bbsp}.
The quark mass parameters giving the optimal fit for all channels
in each family are $m_c=1.51$ GeV,
and $m_b=4.72$ GeV, which agree well with experimental constraints 
\cite{pdg96}. As a consequence of the resummation, we 
observe that in all cases the 
moment ratios are stable as $n$ becomes 
large, in contrast to the conventional results 
obtained by explicitly adding power corrections to a truncated
perturbative series for condensates up
to dimension six or eight. For clarity
a summary of the results presented in the figures is given in 
Tables 1 and 2, where the asymptotic estimates
for the moment ratios are compared with the experimental results
\cite{pdg96} where available. For the vector, axial--vector,
and $A'$ channels the magnitudes of the bound state masses, and also the 
inter--channel splitting, are very well reproduced. For the
scalar and pseudo--scalar channels the magnitudes are reasonably
well approximated but the splitting between these states is not
well described. Possible interpretations for the results
obtained will be presented in the next section. 
\newpage

\begin{table}
\label{resultscc}
\begin{tabular}{||l|l|l|l||} \hline
\hspace{0.2in} {\it current} \hspace{0.2in} 
      & \hspace{0.2in} $M_{expt}$ (GeV) \hspace{0.2in}
      & \hspace{0.2in} $\sqrt{W(\tilde\si)}$ (GeV) \hspace{0.2in}
      & \hspace{0.2in}  $\tilde\si$ (GeV$^2$) \hspace{0.2in} \\ \hline\hline
\hspace{0.4in}$j_V$\hspace{0.4in} &  \hspace{0.4in} 3.10 \hspace{0.3in}
  & \hspace{0.5in} 3.06 \hspace{0.3in} & \hspace{0.35in} 4.6 \hspace{0.3in}\\ 
            \hline
\hspace{0.4in}$j_A$\hspace{0.4in} &  \hspace{0.4in} 3.51 \hspace{0.3in}
  & \hspace{0.5in} 3.52 \hspace{0.3in} & \hspace{0.35in} 4.8 \hspace{0.3in}\\ 
            \hline
\hspace{0.4in}$j_{A'}$\hspace{0.4in} &  \hspace{0.4in} 3.53 \hspace{0.3in}
  & \hspace{0.5in} 3.56 \hspace{0.3in} & \hspace{0.35in} 4.8 \hspace{0.3in}\\ 
            \hline
\hspace{0.4in}$j_S$\hspace{0.4in} &  \hspace{0.4in} 3.42 \hspace{0.3in}
  & \hspace{0.5in} 3.27 \hspace{0.3in} & \hspace{0.35in} 6.0 \hspace{0.3in}\\ 
            \hline
\hspace{0.4in}$j_P$\hspace{0.4in} &  \hspace{0.4in} 2.98 \hspace{0.3in}
  & \hspace{0.5in} 3.19 \hspace{0.3in} & \hspace{0.35in} 3.9 \hspace{0.3in}\\ 
            \hline
\end{tabular}
\caption{Summary of the estimates obtained for the $c\overline{c}$ bound
         state masses, compared to the experimental values
         \protect\cite{pdg96}. Note that 
         $\tilde\si$ denotes the virtuality scale
         at which the asymptotic result was obtained.}
\end{table}

\begin{table}
\label{resultsbb}
\begin{tabular}{||l|l|l|l||} \hline
\hspace{0.2in} {\it current} \hspace{0.2in} 
      & \hspace{0.2in} $M_{expt}$ (GeV) \hspace{0.2in}
      & \hspace{0.2in} $\sqrt{W(\tilde\si)}$ (GeV) \hspace{0.2in}
      & \hspace{0.2in}  $\tilde\si$ (GeV$^2$) \hspace{0.2in} \\ \hline\hline
\hspace{0.4in}$j_V$\hspace{0.4in} &  \hspace{0.4in} 9.46 \hspace{0.3in}
  & \hspace{0.5in} 9.43 \hspace{0.3in} & \hspace{0.35in} 27 \hspace{0.3in}\\ 
            \hline
\hspace{0.4in}$j_A$\hspace{0.4in} &  \hspace{0.4in} 9.89 \hspace{0.3in}
  & \hspace{0.5in} 10.02 \hspace{0.3in} & \hspace{0.35in} 26 \hspace{0.3in}\\ 
            \hline
\hspace{0.4in}$j_{A'}$\hspace{0.4in} &  \hspace{0.4in} -- \hspace{0.3in}
  & \hspace{0.5in} 10.07 \hspace{0.3in} & \hspace{0.35in} 26 \hspace{0.3in}\\ 
            \hline
\hspace{0.4in}$j_S$\hspace{0.4in} &  \hspace{0.4in} 9.86 \hspace{0.3in}
  & \hspace{0.5in} 9.80 \hspace{0.3in} & \hspace{0.35in} 30 \hspace{0.3in}\\ 
            \hline
\hspace{0.4in}$j_P$\hspace{0.4in} &  \hspace{0.4in} -- \hspace{0.3in}
  & \hspace{0.5in} 9.56 \hspace{0.3in} & \hspace{0.35in} 24 \hspace{0.3in}\\ 
            \hline
\end{tabular}
\caption{Summary of the estimates obtained for the $b\overline{b}$ bound
         state masses, compared to the experimental values
         \protect\cite{pdg96}, with 
         $\tilde\si$ again denoting the virtuality scale
         at which the asymptotic result was obtained.}
\end{table}

\section{Discussion}
In this paper we have presented an explicit application of the 
order-dependent mapping approach to the study of the heavy quark
QCD resonance spectrum.
While it proved convenient to study and modify 
the perturbative asymptotics to ensure
consistency with ambiguities in the OPE power corrections, it 
is clear that subsequent to the conformal mapping any obvious 
split between perturbative (short distance)
and nonperturbative (long-distance) contributions is lost.
Thus, other than by fitting data, it does not appear possible to
extract values for the condensates themselves or even to determine the
dimension of the dominant contribution, in contrast to related 
approaches such as that of Dokshitzer, Marchesini, and Webber \cite{dok96a}. 
This requires a clear and well-defined split between the perturbative
and nonperturbative contributions, a distinction which does not emerge 
naturally within the framework of order-dependent mappings.

Nevertheless, at a purely calculational level, applying this formalism to
the bound state spectrum, without the need for explicit introduction of
condensate parameters, we found quite remarkable agreement with experimental
results for the vector, axial-vector, and $A'$ channels
in both $c\overline{c}$ and $b\overline{b}$ families. Indeed
if we use a more accurate extraction of the coupling
at the $\tau$ scale at $O(a^5)$ \cite{jones97a} the results
are even more impressive for these channels, although for consistency
one should then include the next order perturbative coefficients.

However, the results for the scalar and pseudo--scalar channels,
while again having approximately correct magnitudes, were
unsatisfactory in terms of the relative splitting between the
states. This was not improved by a more accurate extraction
of the coupling. It may be that higher order corrections within this 
approach could resolve this discrepancy.  However, considering the
success in the case of the vectorial channels, it appears likely that there
are contributions to the scalar and pseudo--scalar channels which
are missed by the variational approach, e.g. matrix elements
not associated with vacuum condensates. 
Whether this is due to the particular averaging over 
non-perturbative contributions implicit here,
or more closely related to use of the static quark potential
model for normalisation, is currently under investigation.

\section{Acknowledgements}
The authors would like to express their gratitude to D. J. Broadhurst 
for helpful comments, and supplying a copy of Ref. \cite{broad82}
and to Yu. L. Dokshitzer and C. J. Maxwell for helpful discussions. 
The financial support of A.R. by the Commonwealth Scholarship
Commission in the U.K. and the British Council, and of I.L.S.
by the Royal Society and RFBR (grant 96-02-16126-a) is gratefully acknowledged.
I.L.S. also thanks Prof. T.W.B. Kibble and the Theoretical Physics
Group for their warm hospitality at Imperial College where most of this work
was performed. 

\vfill\eject

\appendix

\section{}
\subsection*{Induced Power Corrections in a $\ph^4_{D=0}$ Model}
In this appendix we use a simple zero dimensional model discussed
by Guida, Konishi, and Suzuki \cite{guida96} to illustrate
concretely the power of order-dependent mappings in generating power
corrections to a perturbative series, and consequently producing
a convergent expansion in a situation where the corresponding 
perturbative series is non-Borel summable.

Consider the zero dimensional $\ph^4_{D=0}$ model with partition function
\be
 Z(g) & = & \int_{-\infty}^{\infty} d\ph e^{-\ph^2-g\ph^4},
\ee
and analytically continue the result for $g\in ${\bf C}. With the change
of variables $z=\sqrt{2g}\ph^2$, this is represented in terms of a parabolic
cylinder function $D_{\nu}(a)$ \cite{gr80},
\be
 Z(g) & = & \left(\frac{\pi^2}{2g}\right)^{1/4}e^{\frac{1}{8g}}D_{-1/2}
         \left(\frac{1}{\sqrt{2g}}\right).
\ee

An $N^{th}$ order approximant to this partition function in the
order-dependent mapping may be written in the form
\be
 Z_N(a) & = & (1-a)^{1/2}\ps_N(a,C_N),
\ee
where $\ps_N$ is the $N^{th}$--order truncation of the Taylor series for
a function $\ps$. The order-dependent mapping between the coupling $g$
and the parameter $a$ is given by
\be
 g=\frac{a}{C(1-a)^2}, \label{rel}
\ee
where this relation holds for complex $g$.
Restricted to the positive real axis 
this $N^{th}$ order approximant may be written explicitly as
\be
 Z_N(a) & = & \sum_{k=0}^N\sum_{n=0}^k\frac{\om_n}{C_N^n}(1-a)^{1/2}a^k
            \frac{\Ga(n+k+1/2)}{\Ga(2n+1/2)\Ga(k-n+1)}, \label{zr}
\ee
where $\om_n$ are the perturbative expansion coefficients

Convergence of the sequence of approximants,
\be
 \{Z_N(a,C_N)\} & \stackrel{N\rightarrow\infty}{\longrightarrow} &
             Z(g)
\ee
has been proven in \cite{guida96} for the
analytically continued model, and the details will not be
recalled here. The importance of this result in the present
context for investigating induced power corrections
follows from the structure of the asymptotic series for
$Z(g)$ for $|g|\ll 1$. This expansion has the form
\be
 Z(g) & \sim & \sum_{n=0}^{\infty} \frac{\Ga(2n+1/2)}{n!}
             ((-1)^n+ mi\sqrt{2}e^{1/(4g)})g^n, \label{wc}
\ee 
with $m=\pm 1$ for $\mp 5\pi/2<Arg(g) < \mp \pi/2$ and $m=0$ otherwise. 
If we consider the structure of this expansion for $g=e^{i\pi}g_R$ with
$g_R$ real and positive, then the first term is the standard 
perturbative series about $g_R=0$, which
at large orders has coefficients exhibiting the factorial
growth ($~\sqrt{e/2\pi}\Ga(n)2^{2n}$) associated with 
a Borel non-summable series.
The second term can be interpreted as a complex nonperturbative
``power correction'' accounting for  the imaginary part introduced in
resolving the perturbative ambiguity (c.f. \cite{grunberg94}).

This may be seen more explicitly by performing a Borel transform on the
perturbative series for $Z(e^{i\pi}g_R)$. From
\be
 Z^{pert}(e^{i\pi}g_R) & = & \sum_{n=0}^{\infty}\frac{g_R^n}{n!}\Ga(2n+1/2), 
    \label{pert}
\ee
representing the $\Ga$--function as an integral and transforming
variables, we have formally
\be
 Z^{pert}(e^{i\pi}g_R) & = & \sqrt{\frac{2}{g_R}}\int_0^{\Lambda}db 
                         e^{-b/g_R} B(b),
\ee
where
\be
 B(b) & = & \frac{1}{\sqrt{1-4b}\sqrt{1-\sqrt{1-4b}}},
\ee
exhibiting branch points at $b=0$ and $b=1/4$. Strictly, the transformation
of variables we have performed is only valid for $b\leq 1/4$ and some 
regularization is required to integrate beyond $\La=1/4$. Nonetheless, the 
position of the branch point at $1/4$ can also be inferred directly from the
structure of the large order coefficients $\sqrt{e/2\pi}\Ga(n)2^{2n}$. 

The branch point at $b=0$ may be removed by renormalisation, while
the ambiguity associated with regularising the integral at the
branch point $b=1/4$ is given by the residue as (c.f. the second term
in (\ref{wc}))
\be
 Res(Z^{pert}_{b=1/4}) & = & \sqrt{\frac{2}{g_R}}e^{-\frac{1}{4g_R}}.
\ee
For illustration, we may symbolically represent this
ambiguity as it would appear in a field theory with
\be
 g_R(Q^2) & = & \frac{1}{b_0\ln(Q^2/\La^2)},
\ee
where $\beta(g_R)=-b_0g_R^2+\cdots$. The ambiguity then takes the form
\be
  \exp\left(-\frac{1}{4g_R}\right) & \rightarrow & 
         \left(\frac{\La^2}{Q^2}\right)^{b_0/4},
\ee
which we may loosely interpret as an OPE type ``power correction''.

Thus we can conclude that the weak coupling expansion is only consistent when
one includes a power correction of this form to the perturbative series.
More generally the regularisation of the Borel singularity will also
result in an imaginary contribution (e.g. by passing the contour
above or below the branch point) counteracted by the imaginary part
in (\ref{wc}).

Therefore at least in this rather formal example, the proven convergence of
the order dependent mapping ensures that it can generate a power correction of
this form. Note that it is likely that one can perturb this result
away from the branch cut by considering $Z(e^{i(\pi-\ep)}g_R)$. The imaginary
part of the weak coupling expansion still exists for any 
$\ep<\pi/2$, so one might
expect that a power correction is still required in this case.

\newpage
\section{}
\subsection*{Perturbative Wilson Coefficients}
For completeness, and also to fix our notation, in this appendix we list 
the perturbative Wilson coefficients for the relevant correlators. 
The currents considered are listed below in the format $j_{\Ga}=...(J^{PC})$:
\begin{itemize}

\item {\bf scalar} : $j_S=\overline{\psi}_i\psi_j$ ($0^{++}$) 
\item {\bf pseudo--scalar} : $j_P=i\overline{\psi}_i\gamma_5\psi_j$ ($0^{-+}$) 
\item {\bf vector} : $j_V=\overline{\psi}_i\gamma_{\mu}\psi_j$ ($1^{--}$)
\item {\bf axial vector} : $j_A=(q_{\mu}q_{\nu}/q^2-g_{\mu\nu})
     \overline{\psi}_i\gamma_{\nu}\gamma_5\psi_j$  ($1^{++}$)
\item {\bf A$'$} : $j_{A'}=\overline{\psi}_i\partial_{\mu}\gamma_5\psi_j$ 
                         ($1^{+-}$)
\end{itemize}
 
Using the notation of \cite{rry85} we introduce the following generic
components for the correlators,
\begin{eqnarray}
 A(u) & = & (1+u^2)\left[\frac{\pi^2}{6}+\ln \frac{1+u}{1-u}\ln\frac{1+u}{2}
     +2l\left(\frac{1-u}{1+u}\right)+2l\left(\frac{1+u}{2}\right) \right. 
            \nonumber\\
  &  & \;\;\;\;\;\;\;\;\;\;\;
    \left.   -2l\left(\frac{1-u}{2}\right)-4l(u)+l(u^2)\right]
     +3u\ln\frac{1-u^2}{4u}-u\ln u \\
 A'(u) & = & (1+u^2)\left[2l\left[\left(\frac{1-u}{1+u}\right)^2\right]
     -2l\left(\frac{u-1}{u+1}\right) \right.\nonumber\\
   &  & \;\;\;\;\;\;\;\;\;\;\;\;\;\;\;\;\;\;\;\;
       \left.-3\ln\frac{1-u}{1+u}\ln\frac{1+u}{2}
     +2\ln\frac{1-u}{1+u}\ln u\right],
\end{eqnarray}
where 
\begin{eqnarray}
 l(x) & = & -\int_0^x dt \, \frac{1}{t} \ln (1-t),
\end{eqnarray}
is the Spence function.
Then following the notation of Section 3, the relevant formulae
for each current under consideration are given by:

{\bf Vector Current \protect\cite{rry85} ($\Gamma=\gamma_{\mu}$)}
\begin{eqnarray}
 N_V & = & 0 \\
 \Pi^0_V & = & \frac{1}{2}u(3-u^2) \\
 \Pi^1_V & = & 2\left[\left(1-\frac{u^2}{3}\right)A(u) 
        + P_V(u)\ln\frac{1+u}{1-u} + Q_V(u)\right] \\
 P_V(u) & = & \frac{1}{24}(33+22u^2-7u^4) \\
 Q_V(u) & = & \frac{1}{4}(5u-3u^3). 
\end{eqnarray}

\vfill\eject

{\bf Axial-vector Current \protect\cite{rry85} 
            ($\Gamma=\gamma_{5}\gamma_{\nu}(q_{\mu}q_{\nu}/q^2-g_{\mu\nu})$)}
\begin{eqnarray}
 N_A & = & 1 \\
 \Pi^0_A & = & u^3 \\
 \Pi^1_A & = & \frac{4}{3}\left[u^2A(u) 
        + P_A(u)\ln\frac{1+u}{1-u} + Q_A(u)\right] \\
 P_A(u) & = & \frac{1}{32}(21+59u^2-19u^4-3u^6) \\
 Q_A(u) & = & \frac{1}{16}(-21u+30u^3+3u^5). 
\end{eqnarray}

{\bf {$A'$ Current \protect\cite{rry85} 
               ($\Gamma=\partial_{\mu}\gamma_{5}$)}}
\begin{eqnarray}
 N_{A'} & = & 2 \\
 \Pi^0_{A'} & = & \frac{1}{2}u^3 \\
 \Pi^1_{A'} & = & \frac{2}{3}\left[u^2A(u) 
        + P_{A'}(u)\ln\frac{1+u}{1-u} + Q_{A'}(u)\right] \\
 P_{A'}(u) & = & \frac{1}{16}(13+28u^2+17u^4-2u^6) \\
 Q_{A'}(u) & = & \frac{1}{24}(-39u+47u^3+6u^5). 
\end{eqnarray}

{\bf Scalar Current \protect\cite{broad82} ($\Gamma=${\bf 1})}
\begin{eqnarray}
 N_S & = & 1 \\
 \Pi^0_S & = & \frac{3}{2}u^3 \\
 \Pi^1_S & = & 2\left[u^2A'(u) 
        + P_S(u)\ln\frac{1+u}{1-u} + Q_S(u)\right] \\
 P_S(u) & = & \frac{1}{16}(3+34u^2-13u^4) \\
 Q_S(u) & = & \frac{1}{8}(21u-3u^3). 
\end{eqnarray}
{\bf Pseudo-scalar Current \protect\cite{broad82} ($\Gamma=\gamma_5$)}
\begin{eqnarray}
 N_P & = & 1 \\
 \Pi^0_P & = & \frac{3}{2}u \\
 \Pi^1_P & = & 2\left[A'(u) 
        + P_P(u)\ln\frac{1+u}{1-u} + Q_P(u)\right] \\
 P_P(u) & = & \frac{1}{16}(19-48u+2u^2+3u^4) \\
 Q_P(u) & = & \frac{1}{8}(21u-3u^3). 
\end{eqnarray}

\newpage
\section{}
\subsection*{Resummation of Coulomb Singularities}
In this appendix we justify the resummation procedure used in
Eq.~(\ref{m1}) to avoid the adverse effects of threshold Coulomb singularities
in the moment ratios for large $n$.

As is well known \cite{coulomb} (see also \cite{resum,resum2}), 
the perturbative series for Im$\Pi(s)$
is generically an expansion in powers of $\la/u$, where $u$ is the
quark velocity introduced in Section~3. The series suffers from threshold
singularities for small $u$, and this is accentuated in moment space,
where the dominant contribution to the $n^{th}$ moment comes from
the region $u\sim 1/\sqrt{n}$. In other words, for higher order moments
the system appears more and more nonrelativistic, and the perturbative series
is poorly behaved; the expansion parameter is now $\sqrt{n}\la$.

This behaviour is usually accounted for by treating the nonrelativistic
Coulomb system exactly, leading to a resummation, e.g. in the form
of the Sommerfeld-Sakharov factor \cite{SSfactor},
\be
 Im\Pi(X) & \longrightarrow & \frac{X}{1-\exp(-X)} \;\;\;\;\;\;\mbox{with}
          \;\;\;\;\;\; X=\frac{16\pi^2}{3}\frac{\la}{u}. \label{SS}
\ee
Since our conformal mapping procedure makes use of the same perturbative
coefficients, a resummation of this form is still necessary. However,
the modified virtuality distribution introduced in Section~2 actually
allows an alternative resummation to that mentioned above.

In order to illustrate the procedure, we consider a simple model and study
first the way in which the standard perturbative approach breaks down
for large $n$. Consider the $n^{th}$ moment,
\be
 \label{Mn}
 M_n(Y) & = & \int_0^{\infty} d\si \frac{\rh(\si)}{(\si+Y)^{n+1}},
\ee  
where $\si=s-4m^2$ is the virtuality, and $Y=Q^2+4m^2$. Now assume that the
full spectral density $\rh(\si)$ is known and has the form
\be
 \label{rhex}
 \rh(\si) & = & \left(\frac{\sqrt{\si}}{\la+\sqrt{\si}}\right)^p,
\ee
with $p$ a positive integer, and normalized to $\rh(\si,\la=0)=1$. 
A perturbative expansion of $\rh(\si)$
has the form
\be
 \label{rhpe}
 \rh_{pert}(\si) & \sim & 1-p\frac{\la}{\sqrt{\si}} 
     +O\left(\frac{\la}{\sqrt{\si}}\right)^2\;\sim\; 
      1-\frac{p}{2m}\frac{\la}{u}+O\left(\frac{\la}{u}\right)^2,
\ee
and thus exhibits Coulomb singularities for small $u$.

For large $n$ we may rewrite the moments in the 
form\footnote{We assume that $\rh(\si)$ is smooth \cite{resum2}.},
\be
 M_n(Y) & \sim & Y^{-(n+1)}\int_0^{\infty} d\si \rh(\si) 
            \exp\left(-n\frac{\si}{Y}\right)
              \left(1+O\left(\frac{1}{n}\right)\right) ,
\ee
and observe
that the dominant contribution to the integral corresponds to
virtualities in the region $\si \sim Y/n$ (this is just the threshold behaviour
$u\sim 1/\sqrt{n}$ noted earlier). Then, for the exact moments, $M_n^{ex}$,
given by (\ref{Mn}) with the exact spectral density (\ref{rhex}),
the dominant threshold behaviour at small $u$ (or large $n$)
corresponds to
\be
 \rh_{ex} & \longrightarrow & \left(1+\sqrt{n}\frac{\la}{\sqrt{Y}}\right)^{-p},
\ee
which is positive and finite for all $n$.

If instead we consider the $O(\la)$ perturbative contribution only, we
find
\be
 \rh_{pert} & \longrightarrow & 1-p\sqrt{n}\frac{\la}{\sqrt{Y}},
\ee
which implies that the approximation must break down, and hence
the moments will become unreliable, for large $n$
when $\rh_{pert}$ passes through zero and becomes negative.

We can now introduce a resummation which restores the correct
behaviour of the moments for large $n$. Performing the integration
by parts which ensures consistency of the perturbative ambiguities with
the OPE naturally suggests that we subsume higher order corrections to
the spectral function into the denominator of the virtuality distribution.
If we consider this as a formal transformation, we rewrite the moments 
in the form
\be
     \label{Mnr}
 M_n^{re}(Y) & = & \int^{\infty}_0 d\si' \frac{1}{(f(\si')+Y)^{n+1}},
\ee 
where the function $f(\si')$ obeys the implicit equation
\be
 \si' & = & \int_0^{f(\si')} d\si \rh(\si). \label{fsi}
\ee
This representation follows from considering the change of variables
$\si=f(\si')$ and assuming that $\rh(\si)$ satisfies certain conditions
which ensure the limits $f(\infty)=\infty$ and $f(0)=0$.

Although this transformation is exact, in this paper we work to 
$O(\la)$ and, solving (\ref{fsi}) to this order, we find
\be
 f(\si') & = & \si'+2p\la\sqrt{\si'}.
\ee
Note that at this order, the present resummation essentially
coincides with the Sommerfeld-Sakharov factor (\ref{SS}). The corrections
are of $O(\la^2/u^2)$ and are thus neglected.
Inserting the leading order result in (\ref{Mnr}), and representing it
in the standard form (\ref{Mn}) with a spectral density $\rh_{re}(\si')$,
one finds that for small $u\sim 1/\sqrt{n}$,
\be
 \rh_{re} & \longrightarrow & 
    \left(1+2\sqrt{n}\frac{\la}{\sqrt{Y}}\right)^{-p}.
\ee
This clearly indicates an approximate recovery of the correct
large $n$ dependence associated with $\rh_{ex}(\si)$, despite using only
the $O(\la)$ terms in $\rh(\si)$. 

To illustrate the success of this resummation we plot moment
ratios $R_n=M_n/M_{n-1}$ using the exact ($R_n^{ex}$), perturbative
($R_n^{pert}$), and resummed ($R_n^{re}$) spectral functions, in Fig.~6.

\begin{figure}
 \centerline{%
   \psfig{file=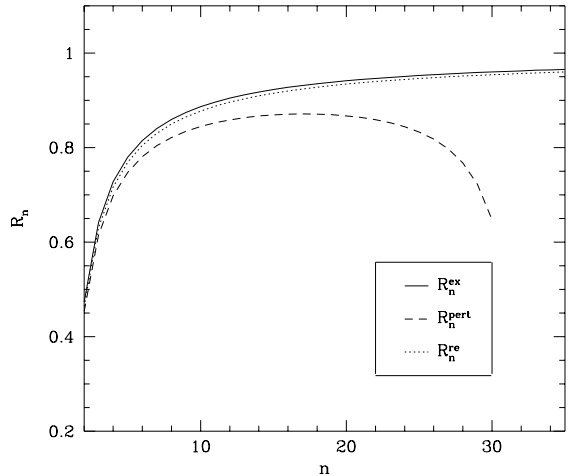,width=8.0cm,angle=0}%
  }
 \caption{Moment ratios are plotted using the exact ($R_n^{ex}$), perturbative
($R_n^{pert}$), and resummed ($R_n^{re}$) spectral functions. We use
the parameters $Y=1$, $\la=0.1$, and $p=1$. Other choices lead to similar
qualitative results.} 
\end{figure}

\noindent One clearly observes the close agreement between $R_n^{ex}$
and $R_n^{re}$ as $n$ becomes large, while the perturbative 
approximation without any resummation breaks down for large $n$.

Furthermore, it is important to note that in the 
resummed case $u\sim 1/\sqrt{n}$
is {\it not} the dominant contribution to the integral. This representation
allows a saddle point to appear for a finite value of $\si$ which
is independent of $n$.
This is associated with the shift in the peak of the virtuality distribution
observed in Fig.~1 for the massless case. Therefore, in practice the 
dominant contribution to the moments does not move closer to the threshold 
region, and as $n$ becomes large the saddle
point dominates the moments, leading to a finite limit. This behaviour
is also observed in the QCD discussion presented in Section~3.

\bibliographystyle{prsty}

\end{document}